\documentclass[pdflatex,sn-mathphys]{sn-jnl}

\usepackage{etoolbox}
\makeatletter
\patchcmd{\ps@headings}
{\hbox to \hsize{\hfill Springer Nature 2021 \LaTeX\ template\hfill}}
{\hbox to \hsize{\hfill Based on  Springer Nature \LaTeX\ template\hfill}}
{}
{}
\patchcmd{\ps@titlepage}
{\hbox to \hsize{\hfill Springer Nature 2021 \LaTeX\ template\hfill}}
{\hbox to \hsize{\hfill Based on  Springer Nature \LaTeX\ template\hfill}}
{}
{}
\makeatother
    
\usepackage{mathtools}
\usepackage{amsmath}
\usepackage{amsfonts}
\usepackage{amssymb}
\usepackage[numbers,sort&compress]{natbib}

\jyear{2022}%

\theoremstyle{thmstyleone}%
\theoremstyle{thmstyletwo}%

\theoremstyle{thmstylethree}%

\begin{document}

\title[Lyapunov Exponents for Temporal Networks]{Lyapunov Exponents for Temporal Networks} 


\author[1]{Annalisa Caligiuri}
\author[1]{Victor M. Eguíluz} 
\author[2]{Leonardo di Gaetano} 
\author[1]{Tobias Galla}
\author*[1]{Lucas Lacasa}\email{lucas@ifisc.uib-csic.es}

\affil[1]{\small{\orgdiv{Institute for Cross-Disciplinary Physics and Complex Systems (IFISC)}, \orgname{CSIC-UIB}, \orgaddress{\city{Palma de Mallorca}, \country{Spain}}}}
\affil[2]{\small{\orgdiv{Department of Network and Data Science}, \orgname{Central European University}, 1100 \orgaddress{\city{Vienna}, \country{Austria}}}}


\abstract{By interpreting a temporal network as a trajectory of a latent graph dynamical system, we introduce the concept of dynamical instability of a temporal network, and construct a measure to estimate the network Maximum Lyapunov Exponent (nMLE) of a temporal network trajectory.  Extending conventional algorithmic methods from nonlinear time-series analysis to networks, we show how to quantify sensitive dependence on initial conditions, and estimate the nMLE directly from a single network trajectory. We validate our method for a range of synthetic generative network models displaying low and high dimensional chaos, and finally discuss potential applications.} 




\maketitle

\section{Introduction}
\label{section:introduction}
Temporal networks (TNs) \cite{TN0, TNX, TN} are graphs whose topology changes in time. They are minimal mathematical models that encapsulate how the interaction architecture of elements in a complex system changes dynamically. TNs have been successfully used 
in a variety of areas ranging from epidemic spreading \cite{TN_epi} or air transport \cite{zanin} to  neuroscience \cite{BrainTN} to cite a few, and it has been shown that important dynamical processes running on networks (e.g. epidemics \cite{TN_epi}, synchronization, search \cite{perra}) display qualitatively different emergent patterns when the substrate is a TN, compared to the case of a static network. These effects are particularly relevant when the timescale of the dynamics running on the graph is comparable to that of the intrinsic evolution of the network, i.e., when there is no manifest separation of timescales. Relatively lesser work has, however, considered the intrinsic dynamics of the network from a principled point of view. Recently, a research programme has been proposed \cite{ACF} in which TNs are to be interpreted as the trajectories of a latent `graph dynamical system' (GDS). The GDS  provides an explicit model for the time-evolution of the network. Similar to a conventional dynamical system (whose output is a time series of scalar or vector quantities), the output of a GDS is a time series of networks, i.e. a TN.\\ 

 The dynamics of TNs and GDS are indeed the objects of ongoing research. For instance, in \cite{ACF} the authors consider how to extend the autocorrelation function of a signal to a graph-theoretical setting. They explore how TNs can oscillate and how harmonic modes, as well as decaying linear temporal correlations of various shapes, emerge. In a similar fashion, the memory of a temporal network has been studied from different angles, including the concept of memory shape \cite{memory}  as a multidimensional extension of memory (high order Markov chain theory) in conventional time series. In this work, we further pursue the abovementioned research framework programme, and consider the problem of dynamical instability and chaos quantification in TNs. Interpreting temporal networks as trajectories in graph space, we aim to generalise the concept of Lyapunov exponents as quantifiers of the sensitivity to initial conditions. Our objective is to define and measure Lyapunov exponents and sensitive dependence to initial conditions of the network as a whole. Our approach is therefore not to quantify chaos in the dynamics for example of every link, but rather to quantify chaos for the collective dynamics of the whole network.\\ 

Since TNs in applications are frequently observed empirically, we focus our implementation and inference of network Lyapunov exponents solely on the observation of a single (long) TN trajectory, without the need to access the underlying GDS (but of course, the framework is also applicable if the GDS is explicitly accessible). Our algorithmic implementation can thus be seen as a conceptual network generalization of the classical algorithms by Wolf and Kantz \cite{wolf, kantz, rosenstein, NTSA}, originally proposed to quantify sensitive dependence on initial conditions directly from empirically observed time series (see also \cite{kurths} for a similarly seminal work).\\

Importantly, any new method needs to be validated. In our case, this is not trivial, since the notion of chaotic TNs is not common in existing literature. A second objective of this work is thus to propose synthetic generative models of chaotic TNs, which can be used as templates to validate the methods we develop for the quantification of chaos in TNs. These methods, once validated, can then be used in wider applications and further research.\\

The rest of the paper is organised as follows. In Section~\ref{section:method} we introduce the theoretical background to our work, and we set the notation. We derive the network analog of the maximum Lyapunov exponent (MLE), and we outline an algorithmic implementation to estimate quantities  such as the spectrum of local expansion rates, trajectory-averaged and volume-averaged expansion rates, and the network maximum Lyapunov exponent (nMLE). In Section~\ref{section:iid} we consider the relatively simple case of random network dynamics as a first example, and show how the method works in  such scenario. Then, in Section~\ref{section:validation_lowdim} we introduce a generative model of (low-dimensional) chaotic temporal networks. This model provides us with `ground-truth' access to the nMLEs of the network trajectories that the model generates. We show that the method we propose to infer nMLEs from a trajectory of networks correctly reconstructs this `true' exponent. We also assess how the estimation of the nMLE is affected when the chaotic network trajectory is polluted with certain amounts of noise, and discuss at this point how to estimate negative nMLEs as well. In Section~\ref{section:validation_highdim} we introduce a different generative model of (high-dimensional) chaotic TNs. We demonstrate that the generated TNs show sensitive dependence to initial conditions, and that the nMLE varies as expected as a function of a network's coupling parameter. In Section~\ref{section:conclusion} we finally conclude and discuss potential applications of the method.

\section{Theory and method}
\label{section:method}
\subsection{Lyapunov exponents for graph dynamical systems}
In nonlinear time series analysis, the maximum Lyapunov exponent $\lambda_{\text{MLE}}$ of a dynamical system quantifies how two trajectories that are initially close separate over time. More precisely, one imagines two copies of the system, which are started from initial conditions at time $t=0$ which are a distance $d_0$ apart. One then defines
\begin{equation}
    \lambda_{\text{MLE}} =\lim_{t\to \infty} \lim_{{d_0\to 0}} \frac{1}{t}\ln \frac{d_t}{d_0}, 
    \label{eq:lyap}
\end{equation}
where $d_t$ is the distance between the two copies of the system at time $t$.

\noindent In practice, the distance $d_0$ is often small but finite, and the limit $d_0\to 0$ may not be accessible. It then turns out that the long-time limit $t\to \infty$ is not accessible either as the growth of $d_t$ is bounded by the size of the attractor of the system \cite{ruelle}. In such cases, the behaviour of $d_t$ usually undergoes a cross-over (at a time which we label $\tau$), between an exponentially expanding phase ($t<\tau$) to a saturated phase ($t>\tau$). In the latter regime, $d_t$ fluctuates around the attractor's size. We will call $\tau$ the saturation time. \\

In the network setting, we assume there exists a (sometimes unknown) graph dynamical system  that determines the evolution of a graph over time. We focus on discrete time. The GDS is then a map, which determines how one network evolves in the next time step. For simplicity, we assume that the set of vertices is fixed, and that the vertices are distinguishable from one another and labelled. Thus, only the set of edges between these vertices evolves in time. A trajectory of the GDS then  consists of a sequence of network snapshots. These trajectories define the TNs generated by the system \cite{TN0, TN}. Each trajectory is given by the sequence of adjacency matrices ${\cal S}=(A_t)_{t\ge0}, $ $A_t=\{a_{ij;t}\}$; $i,j=1,2,\dots,n$ where $a_{ij;t}=1$ if the vertices $i$ and $j$ are connected at time $t$, and zero otherwise (this symmetric choice is for simple, undirected networks, but the method works essentially along the same lines for non-simple and directed networks, perhaps except for different normalization  factors in the definition of the distance, see below). As such, we are considering labelled, unweighted networks of $n$ nodes.\\

It is not clear a priori if the specific choice of the distance used to quantify the deviation between two originally close network trajectories is critical. We conjecture that, as long as this distance is based on the full adjacency matrix --not on a projection of it--, results should hold independent of the specific metric. This is based on the fact that in dynamical systems, the MLE is invariant under different choices of the underlying norm $\lvert\lvert . \rvert \rvert$ \cite{NTSA, ruelle}. There exist many graph distances \cite{graph_distance}. For simplicity, we take an intuitive definition of such a distance which is based on the amount of edge overlap between two networks: given two networks with the same number of nodes $n$ and adjacency matrices  $A$ and $B$, 
\begin{equation}
    d(A,B) = \frac{1}{2}\sum_{i,j} \vert a_{ij} - b_{ij} \vert,
    \label{eq:d}
\end{equation}
where $i$ and $j$ take values from $1$ to $n$\footnote{The pre-factor is used to have a normalized distance when networks are simple, undirected and have the same number of links, otherwise a different normalization is needed.}.  In our setting, networks are simple and unweighted.  In the particular case where $A$ and $B$ have the same number of edges, the distance defined in Eq.~(\ref{eq:d}) is indeed a rewiring distance, i.e., it is given by the number of unique rewirings needed to transform $A$ into $B$, and therefore $d$ is a positive integer-valued function. {One can then further normalize $d$ as appropriate such that it is defined in $[0,1]$, as we will show later. Further details can be found in the Appendix, where we also introduce alternative distances.

\subsection{Inference of network Lyapunov exponents}
\subsubsection{Local expansion rates}
We are interested in quantifying sensitive dependence on initial conditions (and in particular, the network version of $\lambda_{\text{MLE}}$, which we here call $\lambda_{\text{nMLE}}$) when the mechanics of the GDS is not known, and when we only have access to a (single) discrete-time network trajectory ${\cal S}=(A_0,A_1,A_2,\dots)$ of adjacency matrices. This is analogous to the case in which one would like to reconstruct the MLE of a conventional dynamical system from a single time series. The standard approach consists in using Wolf's or Kantz's algorithms \cite{wolf, kantz, rosenstein, NTSA}. The central idea is here to look for recurrences in the orbit, finding points in the time series which might be temporally separated but which are close in phase space. One then monitors the deviation of those points over time. Here, we extend this approach to the case of a time series of networks, i.e. a TN.

We start by fixing an element $A_t$ from $\cal S$, where $t$ is an arbitrary point in time. This adjacency matrix will be the `initial condition' for our analysis. To extend Wolf's algorithm we then proceed to look in $\cal S$ for another element $A_{t'}$ at a different time $t'$, such that $d(A_t,A_{t'})<\epsilon$, where $\epsilon$ is a \textit{small} threshold chosen before the analysis begins\footnote{In practice, $A_{t'}$ is selected as the closest network from $A_{t}$  within the ball of radius $\epsilon$.}. 
This recurrence in phase space allows us to use a single trajectory to explore how two close networks separate over time. We then set $d_0:=d(A_t,A_{t'})$ as the initial distance.
We then proceed to measure how distance evolves over time as we separately track the evolution of $A_{t+k}$ and $A_{t'+k}$ in $\cal S$, where $k=1,2,\dots$. We write $d_k=\lvert\lvert A_{t+k}-A_{t'+k}\rvert\rvert$. Without loss of generality, we can always write the successive distances in terms of a sequence of {\it local expansion rates} $\ell_1, \ell_2, \dots$, 
\begin{equation}
d_k=d_{k-1} \exp(\ell_k).
 \label{eq:seq}
\end{equation}
Each of the $\ell_k$ can be positive (local expansion), negative (local contraction), or zero. Equation~(\ref{eq:seq}) generally models the case where two initially close trajectories ($d_0< \epsilon$) deviate from each other over time. The $\ell_k$ can depend on $k$, since the expansion rates can vary as the trajectories pass through different parts of the attractor \cite{NTSA}. We then define a {\it trajectory-averaged expansion rate}  $\ell$ as follows
\begin{equation}
    \ell = \frac{1}{\tau} \sum_{k=1}^\tau \ell_{k}= \frac{1}{\tau} \sum_{k=1}^\tau \ln \frac{d_k}{d_{k-1}},
    \label{eq:ell}
\end{equation}
where $\tau$ is the saturation time defined earlier. Since we are considering a fixed trajectory (not an ensemble of trajectories), we thus have 
\begin{equation}
\ell=\frac{1}{\tau}\ln \frac{d_{\tau}}{d_0}.
\label{eq:ellfsle}
\end{equation}
Provided the GDS is ergodic (i.e., that a single and long enough orbit adequately visits the whole graph phase space), $\ell$ converges to the network Maximum Lyapunov Exponent $\lambda_{\text{nMLE}}$ in the limit of large $\tau$, independent of the choice of the initial adjacency matrix $A_t$. 
However, in practice, $\tau$ will be finite, and thus we cannot readily assume that $\ell$ fully describes the long-term behaviour, or that it is independent of the initial condition $A_t$. It is thus interpreted as a local Lyapunov exponent \cite{local}, and an average of this quantity over different initial conditions $A_t$ will be required, as discussed further below. 





\begin{figure}[htb]
\begin{centering}
\includegraphics[width=0.95\textwidth]{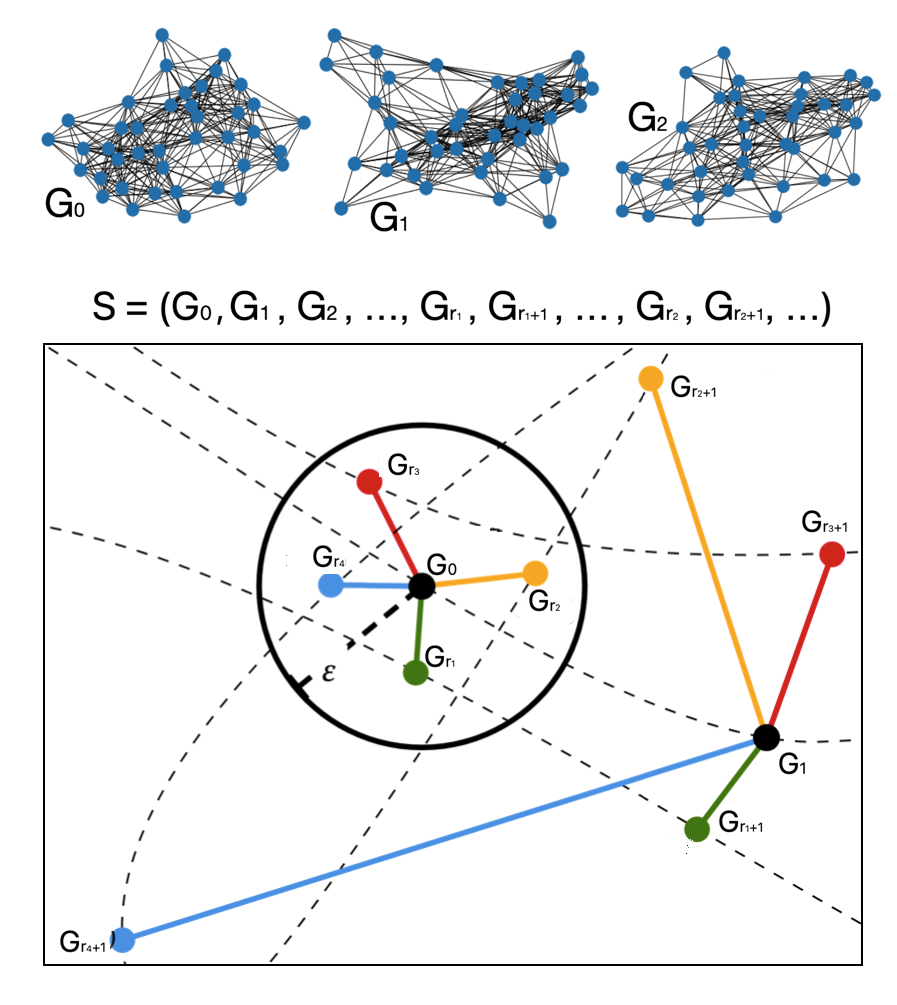}
\par\end{centering}
\caption{Illustration of a temporal network ${\cal S}=(G_0,G_1,G_2,...)$ as a trajectory of a latent graph dynamical system (GDS) and the methodology to compute its network Maximum Lyapunov exponent (Kantz version) $\lambda_{\text{nMLE}}^K$. Each element of the trajectory is a network, i.e. a snapshot of the temporal network. $\lambda_{\text{nMLE}}^K$ is estimated directly from ${\cal S}$ (i.e., without accessing the GDS directly) by looking at recurrences in ${\cal S}$ and quantifying the average expansion around different network snapshots. In this illustration, a ball of radius $\epsilon$ around an arbitrary $G_0$ is fixed, and four recurrences are found where $d(G_0,G_r)<\epsilon$. The initial distance $d(G_0,G_{r})$ is averaged over the four recurrences and the average distance after one time step $d(G_1,G_{r+1})$ is computed. $\lambda_{\text{nMLE}}^K$ is computed by averaging over time, volume and different initial conditions $G_0$ (see the text for details).}
\label{fig:Kantz}
\end{figure}

\subsubsection{Wolf and Kantz methods of estimating maximum Lyapunov exponents for temporal networks}

The time it takes for two trajectories to reach a distance of the order of the attractor's size depends on how close these two trajectories were initially. In other words, the saturation time $\tau$ depends on $d_0$, and therefore on the choice of the threshold $\epsilon$. The limits $d_0\to 0$ and $\tau\to\infty$ are thus related to one another. Conceptually, we would like the trajectories to be as close as possible initially, so that we can monitor the expansion rate for long times, allowing us to obtain a global MLE as opposed to a local Lyapunov exponent. To do this, we need to track the expansion for sufficiently long, but we also need to avoid the regime in which the distance is limited by the characteristic size of the attractor.\\ 

\noindent{\bf Generalised Wolf approach to measuring the nMLE.} We now construct the generalisation of Wolf's approach, which will yield an estimation of the nMLE that we call $\lambda_\text{nMLE}^W$. The aim is to compute $ \langle \ell\rangle=  \langle \frac{1}{\tau}  \ln \frac{d_{\tau}}{d_{0}}\rangle$, where the average is over choices of $A_t$ and $A_{t'}$. In practice one considers a set of $w$ initial choices of $A_t$, which we index $i=1,\dots, w$, and for each of these choices, one additional point $A_{t'}$ on the trajectory such that $d(A_t,A_{t'})<\epsilon$. One then obtains
\begin{equation}
    \lambda_\text{nMLE}^{W} = \frac{1}{w}\sum_{i=1}^w \ell(i),
    \label{eq:wwolf_lyap}
\end{equation}
where $\ell(i)$ is the trajectory-averaged expansion rate $\ell$ computed for the $i$-th initial condition, obtained from Eq.~(\ref{eq:ell}).\\
Algorithmically, this approach has the advantage that we do not need to fix the choice of $\tau$, we are able to flexibly adjust $\tau$ for each initial condition, according to the specific $d_0$ we are able to find in $\cal S$. On the other hand, this approach is point-wise, in the sense that for each choice of $A_t$ one only considers a single $A_{t'}$ nearby. As a consequence, this method does not necessarily capture the {\em average} expansion rate around each initial condition $A_t$.\\

\noindent{\bf Generalised Kantz approach to measuring the NMLE.} In order to calculate such an average expansion rate (for a given choice of $A_t$) one would, for a fixed $A_t$, have to average over the expansion rates for choices of $A_{t'}$ in an $\epsilon$-ball about $A_t$. This {\it volume-averaging} is the basis of Kantz's generalization \cite{kantz, rosenstein} of Wolf's algorithm \cite{wolf}, see Fig.\ref{fig:Kantz} for an illustration. For a given initial condition $A_t$ Kantz' method provides a trajectory and volume averaged expansion rate $ \langle \frac{1}{\tau} \ln \frac{d_{\tau}}{d_{0}} \rangle_{\text{volume}}$. We will write $\Lambda$ for the volume-averaged expansion rate. For fixed $A_t$, this could algorithmically be obtained as follows. One chooses $N$ different $A_{t'}$ from the trajectory ${\cal S}$, all within distance $\epsilon$ from $A_t$. We label these $j=1,\dots,N$. For each $A_{t'}$ one then computes $\ell(j)$ via Eq.~(\ref{eq:ell}). Then one sets \begin{equation}
    \Lambda(A_t) = \frac{1}{N}\sum_{j=1}^N \ell(j), 
    \label{L1}
\end{equation}
where $N$ is the number of initial conditions inside a ball of radius $\epsilon$ and centered at $A_t$ that we have found in the sequence $\cal S$.\\ 
In practice, Kantz algorithm proceeds slightly differently. Instead of first computing the $\ell(j)$, and then averaging the expansion rates, the average over the $A_{t'}$ is instead computed at the level of distances. That is to say, one makes $N$ choices of $A_{t'}$ as described above, and then obtains $d_k(j)$ for each $j=1,\dots, N$ and $k=0,1,2,\dots$ ($k$ runs up to the relevant cut-off time). One then sets
\begin{equation}
    \Lambda(A_t)= \frac{1}{\tau} \ln   \frac{N^{-1}\sum_{j=1}^N d_\tau(j)}{N^{-1}\sum_{j=1}^N d_0(j)},
\label{L2}
\end{equation}
where $\tau$ is a priori fixed for all $j$. The numerator in the logarithm represents the volume average (over choices of $A_{t'}$ in a ball about $A_t$) of $d_\tau$, and the denominator is the volume average of $d_0$.

We stress that Eqs.~(\ref{L1}) and (\ref{L2}) are mathematically different and do not necessarily lead to the same results. While Eq.~(\ref{L1}) allows adapting the precise value $\tau$ (which depends on $d_0$) for each trajectory, Eq.~(\ref{L2}) instead requires us to use a uniform $\tau$ across all trajectories. The latter expression looks at the expansion in time of an initially small volume centered on $A_t$, and is closer in spirit to capturing the underlying nonlinear dynamics than Eqs.~(\ref{L1}). Accordingly, Eq.~(\ref{L2}) is the basis of the Kantz' method. 

The quantity $\Lambda$ in Eq.~(\ref{L2}) is still a local quantity, in the sense that it was computed for a phase space volume around a fixed choice of $A_t$. In principle, the local volume-averaged expansion rate could vary across different regions in phase space. To capture the global long-term behaviour we therefore additionally average over choices of $A_t$, and then finally obtain the global volume-averaged network maximum Lyapunov exponent \begin{equation}
    \lambda_\text{nMLE}^K= \bigg \langle   \frac{1}{\tau} \ln  \frac{\langle d_{\tau}\rangle_{\text{volume}}}{\langle d_{0} \rangle_{\text{volume}}} \bigg \rangle_{A_t},\label{L3}
    \end{equation}
    where we have written $\langle\cdots\rangle_{A_t}$, for the average over initial conditions $A_t$. In practice this is carried out by averaging over a set of $w$  choices of $A_t$, i.e.,
 \begin{equation}
    \lambda_\text{nMLE}^K =   \frac{1}{w} \sum_{j=1}^w \Lambda(A_t^{(j)}) 
    \label{eq:kantz_lyap}
\end{equation}
where $\Lambda(A_t^{(j)})$ stands for the expression in Eq.~(\ref{L2}) for the fixed choice $A_t=A_t^{(j)}$.\\ 

We note that the value of the saturation time $\tau$ or the radius of the ball $\epsilon$ will have to be selected after some numerical exploration. Indeed, a better estimate of the Lyapunov exponent is obtained when the cut-off time $\tau$ is large. This, in turn, is the case when the initial distance between $A_t$ and $A_{t'}$ is small, hence favouring choosing a relatively small value of $\epsilon$. However, a small value of $\epsilon$ complicates the task of finding points $A_{t'}$ that are at most a distance $\epsilon$ away from $A_t$ on the given trajectory. In practice, a trade-off is to be struck.\\


To summarise, from a given TN trajectory (i.e. a sequence of network snapshots) we first measure the local expansion rates $\{\ell_k\}_{k=1}^{\tau}$ via Eq.~(\ref{eq:seq}) for a fixed choice of $A_t$ and $A_{t'}$. The set of $\ell_k$ obtained in this way provide information on the fluctuations of the local expansion rate (for fixed $A_t$ and $A_{t'}$), and its trajectory-average $\ell$. We can then proceed along two alternative routes. In the first approach (i) we average $\ell$ over different choices for the initial condition $A_t$ [Eq.~(\ref{eq:wwolf_lyap})], and obtain Wolf's approximation to the nMLE $\lambda_\text{nMLE}^W$.
Alternatively, (ii) we can initially perform, for each initial condition $A_t$, an average over $A_{t'}$. 
This is done by computing the local expansion of a volume $\langle d_k \rangle_{\text{volume}}$ and then averaging this over time [Eq.~(\ref{L2})].  This is then repeated for different choices of $A_t$, and once hence obtains a distribution $P(\Lambda)$ describing the fluctuations across different points in the network phase space. Its mean provides Kantz's approximation to the nMLE  $\lambda_{\text{nMLE}}^K$.\\

\noindent In the following sections, we present a validation of this method for random, low-dimensional and high-dimensional chaotic temporal networks.

\section{The white-noise equivalent of a temporal network: independent and identically distributed random graphs}
\label{section:iid}
Before addressing the case of chaotic dynamics, we briefly discuss the case of random network trajectories, with no correlations in time. One then expects no systematic expansion or compression in time, and the resulting Lyapunov exponent should hence vanish\footnote{By construction, $d_1>d_0$ as we force $d_0<\epsilon$, so in order to avoid spurious expansions at $k=1$, in this section we don't take into account $d_0$ in the estimation of finite Lyapunov exponents, i.e. our starting time is $k=1$.}. We here seek to verify that this is indeed the case for the procedures we have introduced to estimate the Lyapunov exponents of TNs.
Studying this is of interest, among other reasons, because an empirically obtained time series may appear random. It is then important to be able to decide if the trajectory is consistent with an uncorrelated random trajectory in network space, or with a deterministic chaotic model.\\
Here we study the simple case where $\cal S$ is an  independently drawn sequence of Erd\"os-R\'enyi graphs with $n$ nodes and in which the probability that any two nodes are connected is $p$. This is an analog to white noise for TNs, i.e., a situation in which the TN displays delta-distributed autocorrelation function \cite{ACF}.\\

\noindent At odds with a deterministic GDS, the distances between different points on a network trajectory are now random variables. More precisely, since all elements of $\cal S$ are the adjacency matrices of Erd\"os-R\'enyi graphs, the elements of these matrices are Bernoulli variables, taking values zero (with probability $1-p$) or one (with probability $p$). For independent adjacency matrices $A$ and $B$, the possible values of $\lvert a_{ij} - b_{ij}\rvert$ are then zero with probability $p^2+(1-p)^2$, and one with probability $2p(1-p)$. Thus we have 
\begin{equation}
   d(A,B) \sim \text{Binomial}[n(n-1)/2, 2p(1-p)].
    \label{eq:bin}
\end{equation}
  Eq.~(\ref{eq:bin}) is numerically verified in panel (a) of Fig.~\ref{fig:iid}. \\

\begin{figure}[htb]
\begin{centering}
\includegraphics[width=0.49\textwidth]{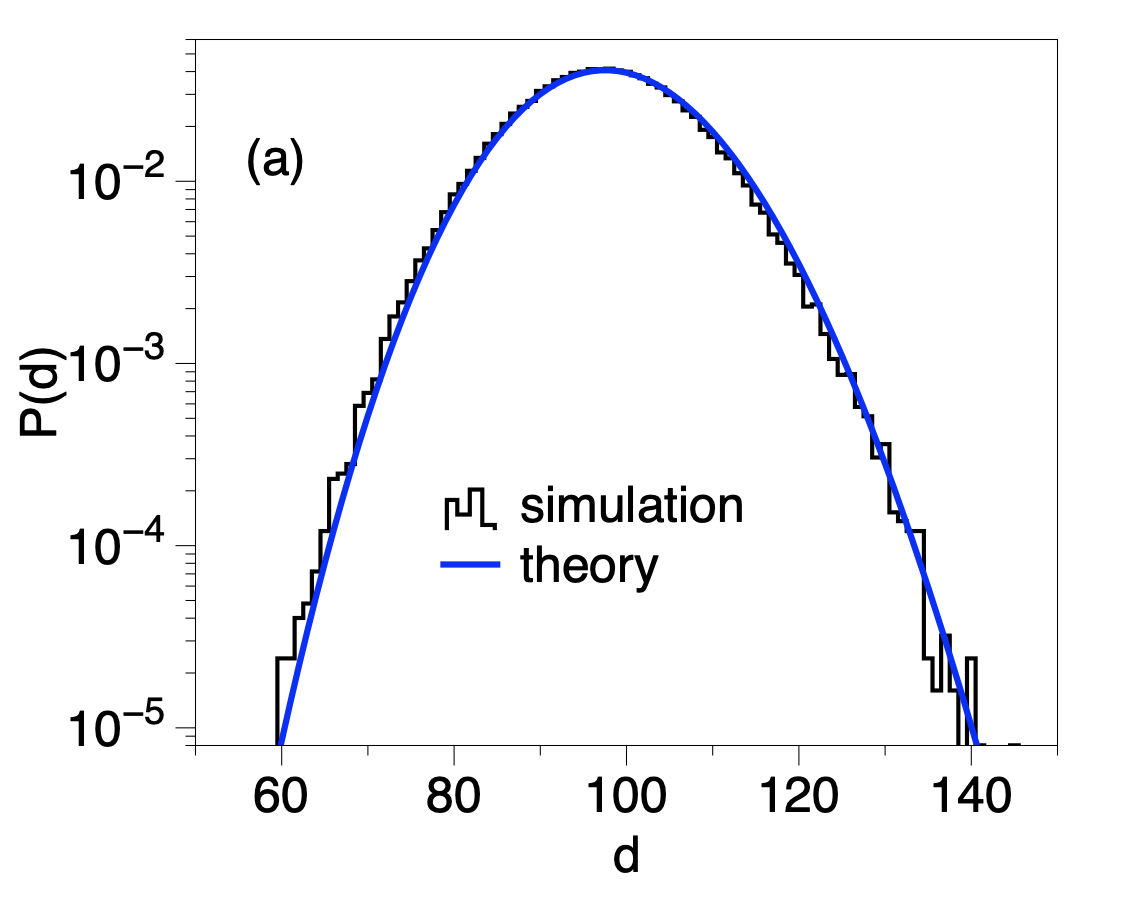}
\includegraphics[width=0.49\textwidth]{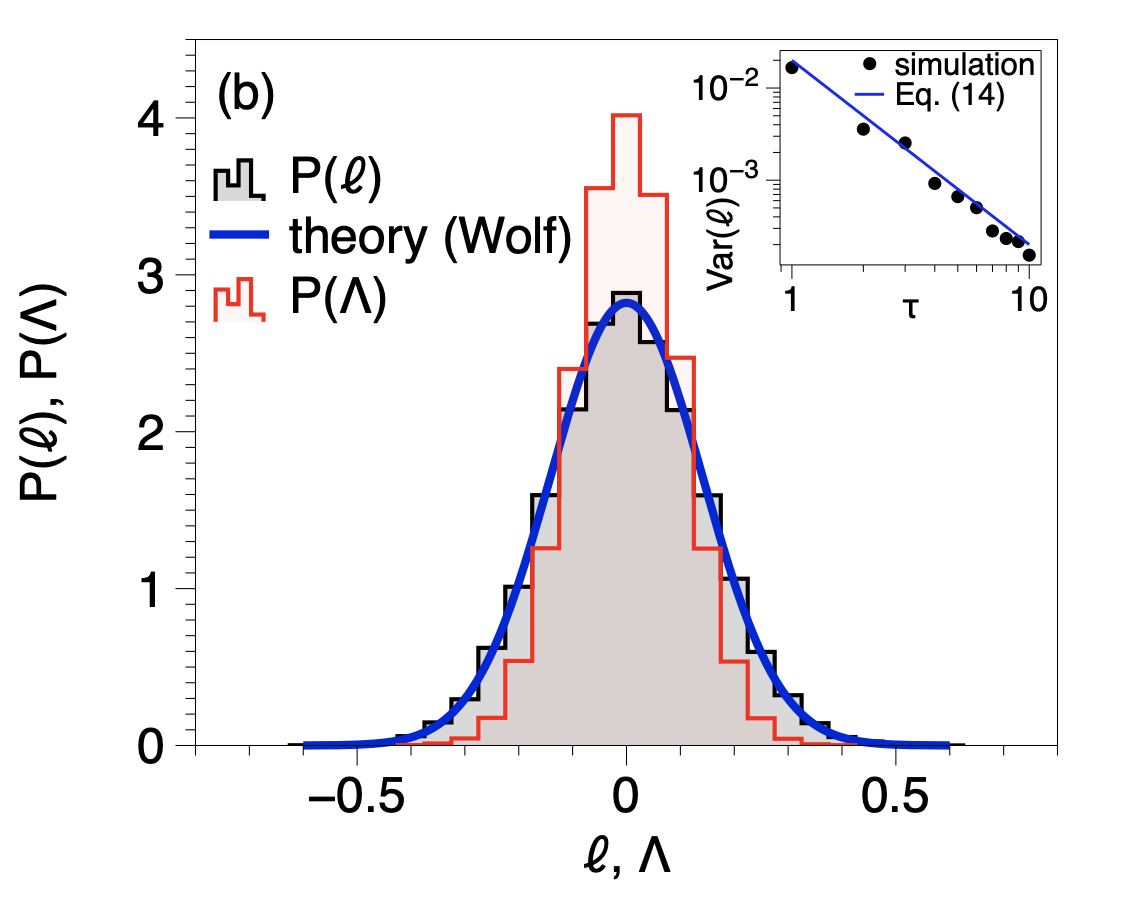}
\par\end{centering}
\caption{
Panel (a): Probability distribution of the distance between consecutive networks in an i.i.d. sequence of $ \vert {\cal S} \vert =10^5$ Erdos-Renyi graphs with $n=100$ nodes and $p=0.01$ (black line). The plot is in semi-log, where a Gaussian shape appears as an inverted parabola, so we can better appreciate the tails. Blue solid line is Eq.~(\ref{eq:bin}). 
 Panel (b): $P(\ell)$  (black) and $P(\Lambda)$ (red) associated with a sequence of $\vert {\cal S}\vert  = 5 \cdot 10^4 $ i.i.d. ER networks ($n=100,p=0.01$), for $\tau=1$ and $w=\mathcal{O}(10^4)$ initial conditions. The mean of both distributions (nMLE) is essentially zero for both methods, but the dispersion around the mean is larger in Wolf's approach. The solid blue line is the theoretical prediction, i.e. a Gaussian distribution with mean $0$ and variance as in Eq.~(\ref{eq:Var}) with $\tau=1$. The inset of panel (b) shows how the variance of $\ell$ shrinks as $\tau$ increases (resulting in a much lower uncertainty around a zero nMLE): dots are different numerical simulations with $ \vert {\cal S} \vert = 10^4$ and $w=10^3$, for different $\tau$; the blue line is Eq.~(\ref{eq:Var}).}
\label{fig:iid}
\end{figure}

The quantity $\ell$ in Eq.~(\ref{eq:ellfsle}) is given by $\ell = \frac{1}{\tau}\left(\ln\,d_\tau - \ln d_0\right)$. As we have just established, $d_0$ and $d_\tau$ are independent binomial random variables following the distribution in Eq.~(\ref{eq:bin}). For large networks ($n\gg 1$) this distribution can be approximated as a Gaussian, with mean $\mu=qn(n-1)/2$ and variance $\sigma^2= q (1-q)n(n-1)/2$, where we have written $q\equiv 2p(1-p)$. Writing $d_0=\mu+\sigma z_0$, with $z_0$ a standard Gaussian random variable, we have 
\begin{equation}
\ln\,d_0\approx\ln\left[\mu(1+\frac{\sigma}{\mu }z_0)\right] = \ln\,\mu + \frac{\sigma}{\mu} z_0 - \frac{1}{2}\frac{\sigma^2}{\mu^2} z_0^2 + \dots,
\end{equation}

after an expansion in powers of $\sigma/\mu$, where the latter quantity is of order ${\cal O}(1/n)$. The same expansion can be carried out for $d_\tau$, and we therefore find 
\begin{equation}
\ell = \frac{1}{\tau}\left[\frac{\sigma}{\mu} (z_\tau-z_0) +\frac{1}{2}\frac{\sigma^2}{\mu^2}(z_0^2-z_\tau^2)\right]+\dots .
\end{equation}
We note that the second term in the bracket is of sub-leading order in $1/n$.
Hence, $\ell$ is to lowest order in $1/n$  approximately Gaussian, with mean zero and variance 
\begin{equation}\label{eq:Var}
\mbox{Var}(\ell)=\frac{2}{\tau^2}\frac{\sigma^2}{\mu^2}=\frac{1}{\tau^2}\frac{4(1-q)}{qn (n-1)}
\end{equation}
This theory has been numerically verified, and in panel (b) of Fig.\ref{fig:iid} we plot $P(\ell)$ both for $\tau=1$ (outer panel) and Eq.\ref{eq:Var} for increasing values of $\tau$ (inset panel).\\
\noindent The case of $\Lambda$ (Kantz-version) should intuitively converge even faster than $\ell$ (Wolf-version) since in this case we are carrying out two averages instead of just one, i.e. $P(\Lambda)$ should have a smaller variance than $P(\ell)$, for a given $\tau$. 
This is confirmed in panel (b) of Fig.~\ref{fig:iid}, where we also observe that both methods yield the same (correct) estimation of the nMLE, which in this case is approximately zero (both estimates are of the order of $10^{-6}$). Note that the main panels of Fig.~\ref{fig:iid}) are for the case $\tau=1$, so it is a worst-case scenario: as $\tau$ increases $\text{Var}(\ell)$ shrinks [Eq.~(\ref{eq:Var})] and the uncertainty around the null shrinks accordingly [see inset of Fig.~\ref{fig:iid}(b)].\\
We were not able to find a closed-form solution for $P(\Lambda)$ as averages inside the $\epsilon$-ball are random variables whose distribution explicitly depends on the specific initial condition $A_t$: this calculation is left as an open problem. In anycase, we conclude that an i.i.d. temporal network has a null MLE Lyapunov exponent and the methodology (in  both variants) correctly estimates it.  


\section{Low-dimensional chaotic networks} 
\label{section:validation_lowdim}
\subsection{Network generation: the dictionary trick}

To be able to validate the method in the context of chaotic dynamics, we ideally need to have access to chaotic network trajectories with a ground true nMLE. This is  difficult as a general theory of chaotic GDS is not yet accessible. To circumvent this drawback, in this section we develop a method to construct (low-dimensional) chaotic network trajectories by symbolising in graph space-time series from low-dimensional chaotic maps. The method of graph-space symbolisation was first proposed as a so-called `dictionary trick' in \cite{ACF} and consists of the following steps: 
\begin{itemize}
    \item We construct a network dictionary ${\cal D}$. This is a set of networks that allows us to map a real-valued scalar $x\in [0,1]$\footnote{We choose the interval $[0,1]$ without loss of generality.} into a network, such that the distance between two scalars is preserved in graph space. The set ${\cal D}$ is therefore ordered and equipped with a metric, such that the distance between two real-valued scalars $\vert x-x'\vert $ is preserved in the graph symbols. More concretely, the dictionary of networks ${\cal D}=(G_1,G_2,...,G_L)$ such that $d(G_p,G_q) \propto \lvert p-q\rvert $ (one can subsequently normalize $d$ according to the length of the dictionary, such that we have $d\in [0,1]$).
     \item Once such dictionary is built, any one-dimensional time series can be mapped into a sequence of networks. In particular, we can map chaotic time series with well-known MLEs into network trajectories, from which an independent estimate of the nMLE can be obtained.
\end{itemize}  

Algorithmically, the dictionary is generated sequentially with $G_1 \sim \text{ER}(p)$ (an Erd\"os-R\'enyi graph with parameter $p$) and then iteratively constructing $G_{k+1}$ from $G_k$ by rewiring a link that (i) has not been rewired in any previous iteration of the algorithm, (ii) into a place that did not have a link in any previous iteration of the  algorithm. It is easy to see that such algorithm ensures that ${\cal D}$ provides a partition of $[0,1]$ of the form $[0,1]=\cup_{k=0}^{L-1} [k/L, (k+1)/L]$, where $L$ is the number of networks in the dictionary. The dictionary is thus metrical, in the sense that the rewiring distance between any two elements in the dictionary is (for a sufficiently large refinement $L$) arbitrarily close to the associated real-valued scalars in the original interval. Once the dictionary is established, we can then generate synthetic temporal network trajectories as symbolizations of unit interval dynamics by matching points in the subinterval $[k/L, (k+1)/L]$ to the symbol $G_{k+1}$. The resulting temporal network $\cal S$ inherits, by construction, the properties of the scalar time series, and in particular can be used to generate chaotic TNs.\\

\begin{figure}[htb]
\begin{centering}
\includegraphics[width=0.45\textwidth]{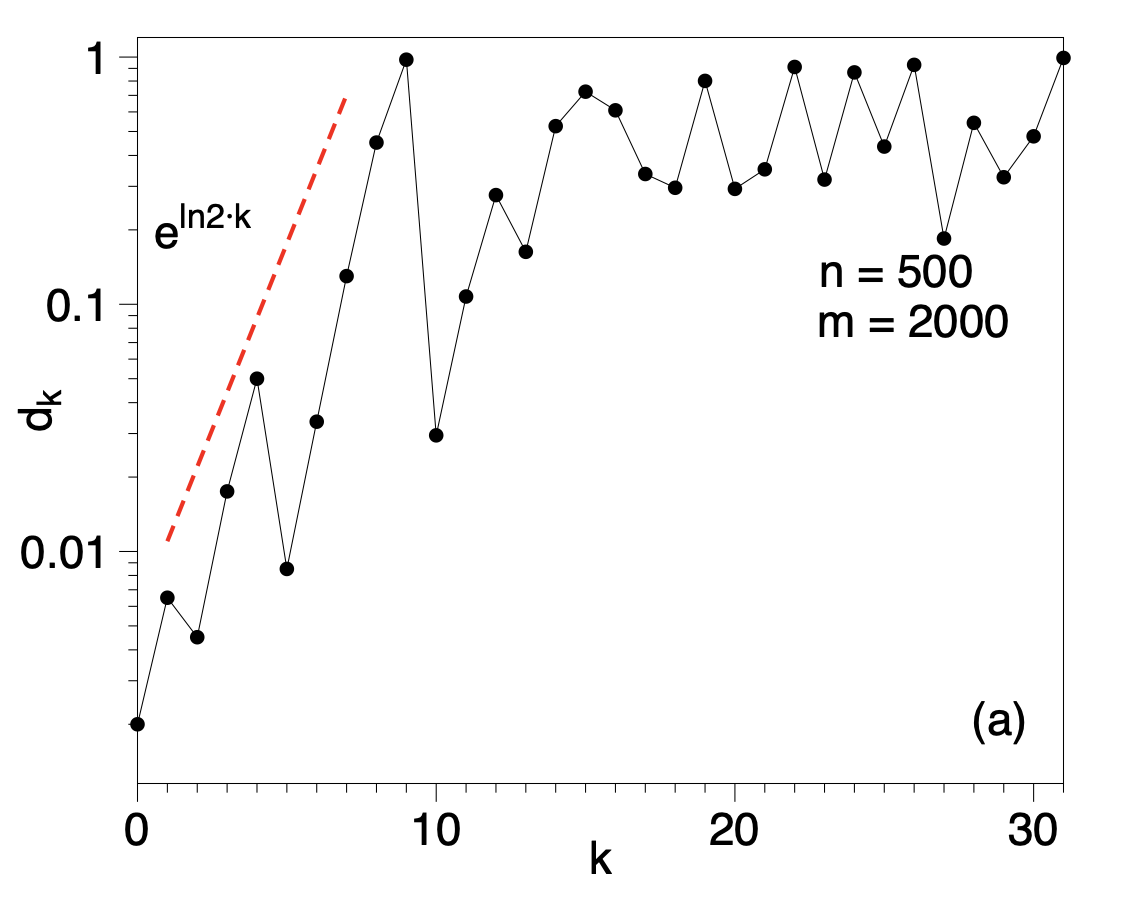}
\includegraphics[width=0.45\textwidth]{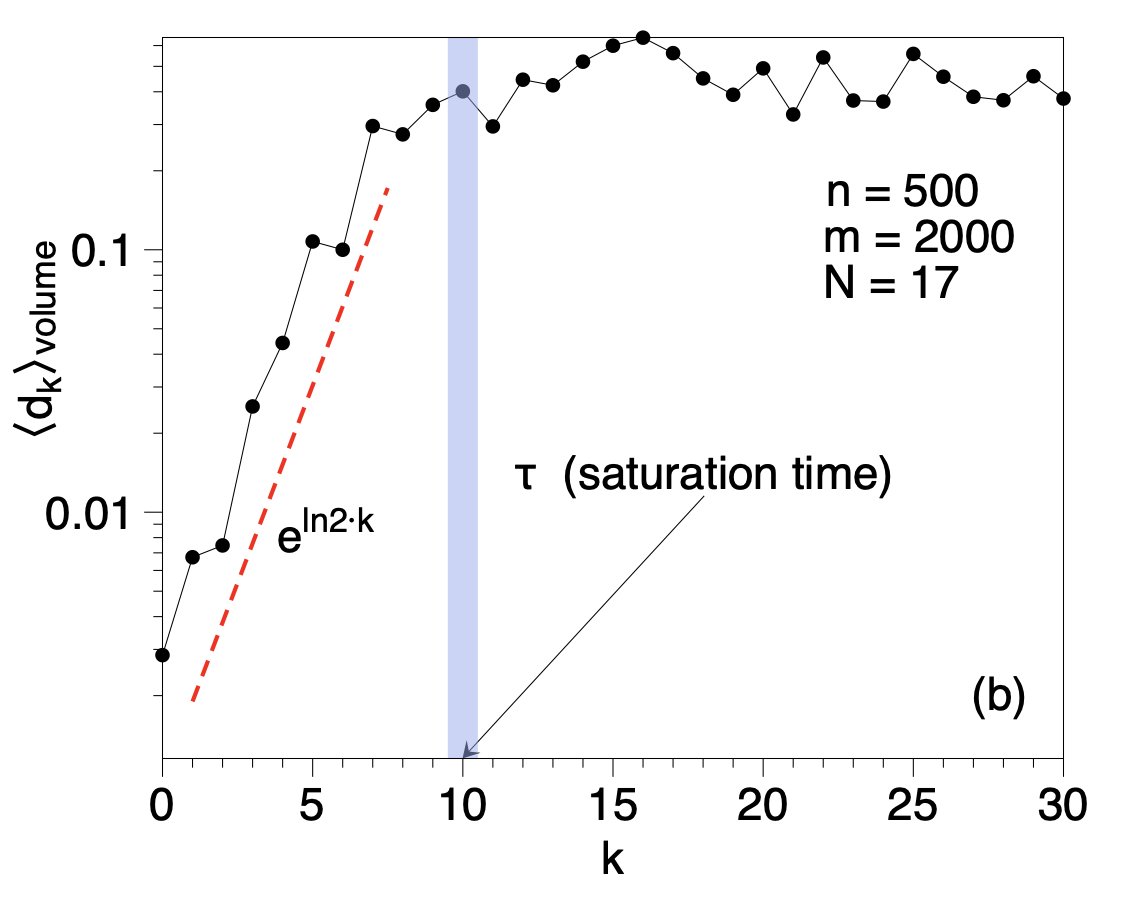}
\par\end{centering}
\caption{Panel (a): Semi-log plot of the distance $d_k$ as a function of iteration index $k$, for two initially close network trajectories sampled from $\cal S$. We can appreciate an initial exponentially expanding phase, followed by a saturation phase, although the local expansion rate strongly fluctuates. Panel (b) Volume-averaged distance $\langle d_k\rangle_{\text{volume}}$ as a function of time $k$, for $N=17$ initial graph conditions inside a volume centered at an initial graph of $n=500$ nodes and $m=2000$ edges. Network dynamics evolve according to a logistic map as described in the text, whose true Lyapunov exponent is $\ln 2\approx 0.693$. We can see how the volume enclosing the graphs on average expands exponentially fast --with an exponent close to $\ln 2$, as expected-- until it reaches the attractor size, what happens at the saturation time $\tau\approx 10$.}
\label{fig:d_vs_t_kantz}
\end{figure}

\subsection{Results for the logistic map}
As a first validation, we consider the fully chaotic logistic map 
\begin{equation}
    x_{t+1}=4x_t(1-x_t),\hspace{0.5cm} x_t\in[0,1],
    \label{eq:log4}
\end{equation}
 that generates chaotic trajectories with $\lambda_{\text{MLE}}=\ln 2\approx 0.693$. Using the dictionary trick, from a signal extracted from Eq.~(\ref{eq:log4}) we generate a temporal network trajectory $\cal S$ of $\vert {\cal S} \vert = 3000$ network snapshots. In this validation, networks have $n=500$ nodes and $m=2000$ edges.\\

\begin{figure}[htb]
\begin{centering}
\includegraphics[width=0.48\textwidth]{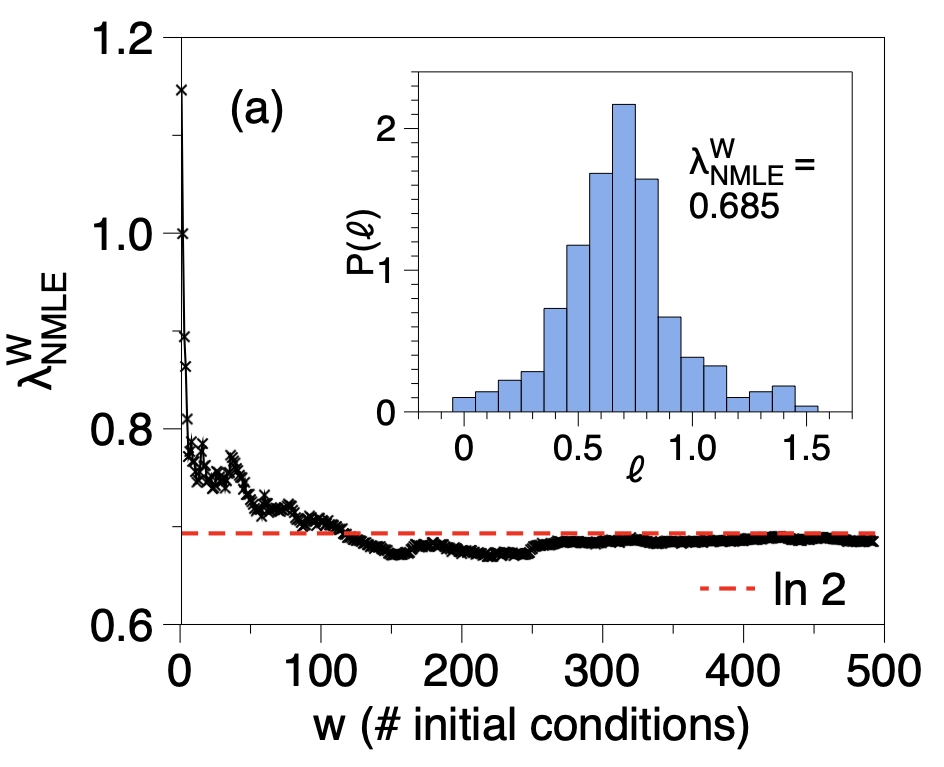}
\includegraphics[width=0.48\textwidth]{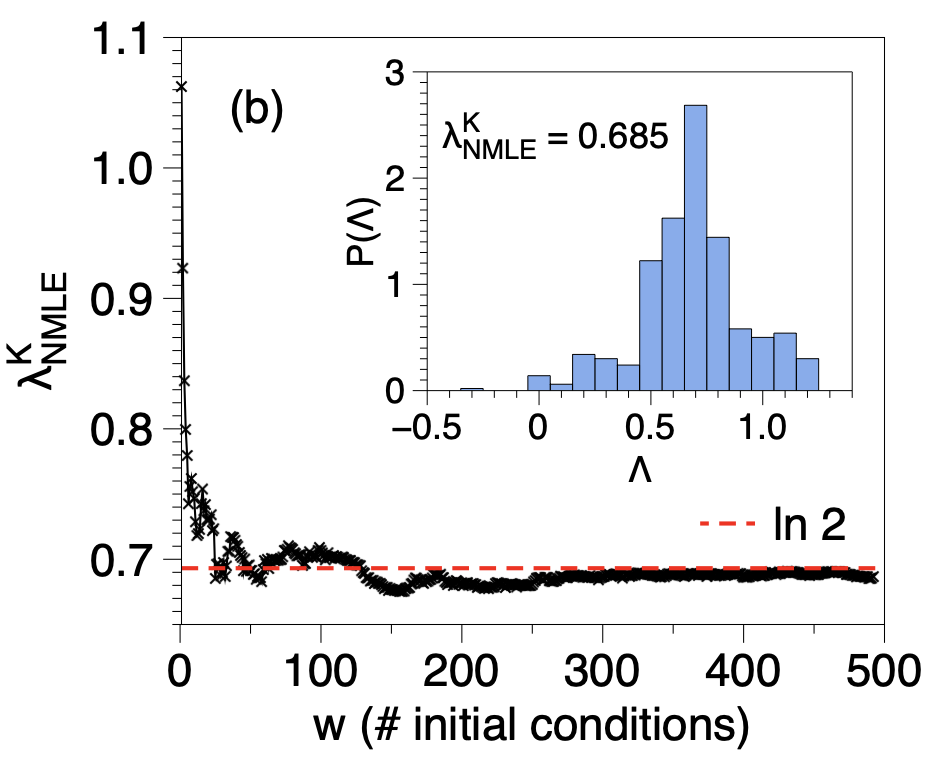}
\par\end{centering}
\caption{Panel (a): Approximation to $\lambda_{\text{nMLE}}^W$ following Wolf's approach (see  text), computed by averaging $\ell$ over $w$ randomly sampled initial conditions [Eq.\ref{eq:wwolf_lyap}], as a function of $w$. We can see that the exponent converges to the ground true exponent $\ln 2$ as $w$ increases. Inset in (a): Probability distribution $P(\ell)$, sampled by estimating $\ell$ for $w=500$ different initial graph conditions sampled randomly from $\cal S$. The mean of this empirical distribution is $\lambda_{\text{nMLE}}^W=\langle \Lambda \rangle_{A_t}\approx 0.685$, very close to the true exponent $\ln 2 \approx 0.693$. Panel (b): Same as panel (a), but using Kantz's approach (see the text), where we compute the volume and trajectory averaged expansion rate $\Lambda$ for $w$ initial conditions. Convergence properties are similar in both cases.}
\label{fig:lyap_kantz}
\end{figure}

 For illustration, in panel (a) of Fig.~\ref{fig:d_vs_t_kantz} we plot in semi-log scales the (properly normalized) distance $d_k$  as a function of the iteration index $k$, for two initially close network trajectories sampled from $\cal S$. We can see an initial exponentially expanding phase (whose exponent is an estimation of $\ell$) followed by a saturation, although the distance function shows strong fluctuations. To cope with these, in panel (b) of the same figure we plot the
volume-averaged expansion $\langle d_k \rangle_{\text{volume}}$ vs $k$ for a ball of radius $\epsilon=0.005$ centered at a specific initial graph from $\cal S$. We can now clearly see the initial exponential phase followed by a cross-over to a saturation phase. The cross-over marks the saturation time $\tau$ where the distance reaches the attractor size. Note that the slope of the exponential expansion (i.e. the estimate of $\Lambda$) is close to $\ln 2$, the true MLE.\\

Figure~\ref{fig:lyap_kantz} shows the estimated of the nMLE obtained both using Wolf's approach [panel (a)] and Kantz's approach [panel (b)]. These are from averaging $\ell$ (Wolf) and $\Lambda$ (Kantz) over $w=500$ initial graph conditions sampled from $\cal S$. In both cases, the average quickly stabilises for $w\approx 100$, and we obtain estimates $\lambda_{\text{nMLE}}^W\approx \lambda_{\text{nMLE}}^K \approx 0.685$, very close to the ground true $\lambda_{\text{MLE}}=\ln 2 \approx 0.693$. 


\begin{figure}[htb]
\begin{centering}
\includegraphics[width=0.6\textwidth]{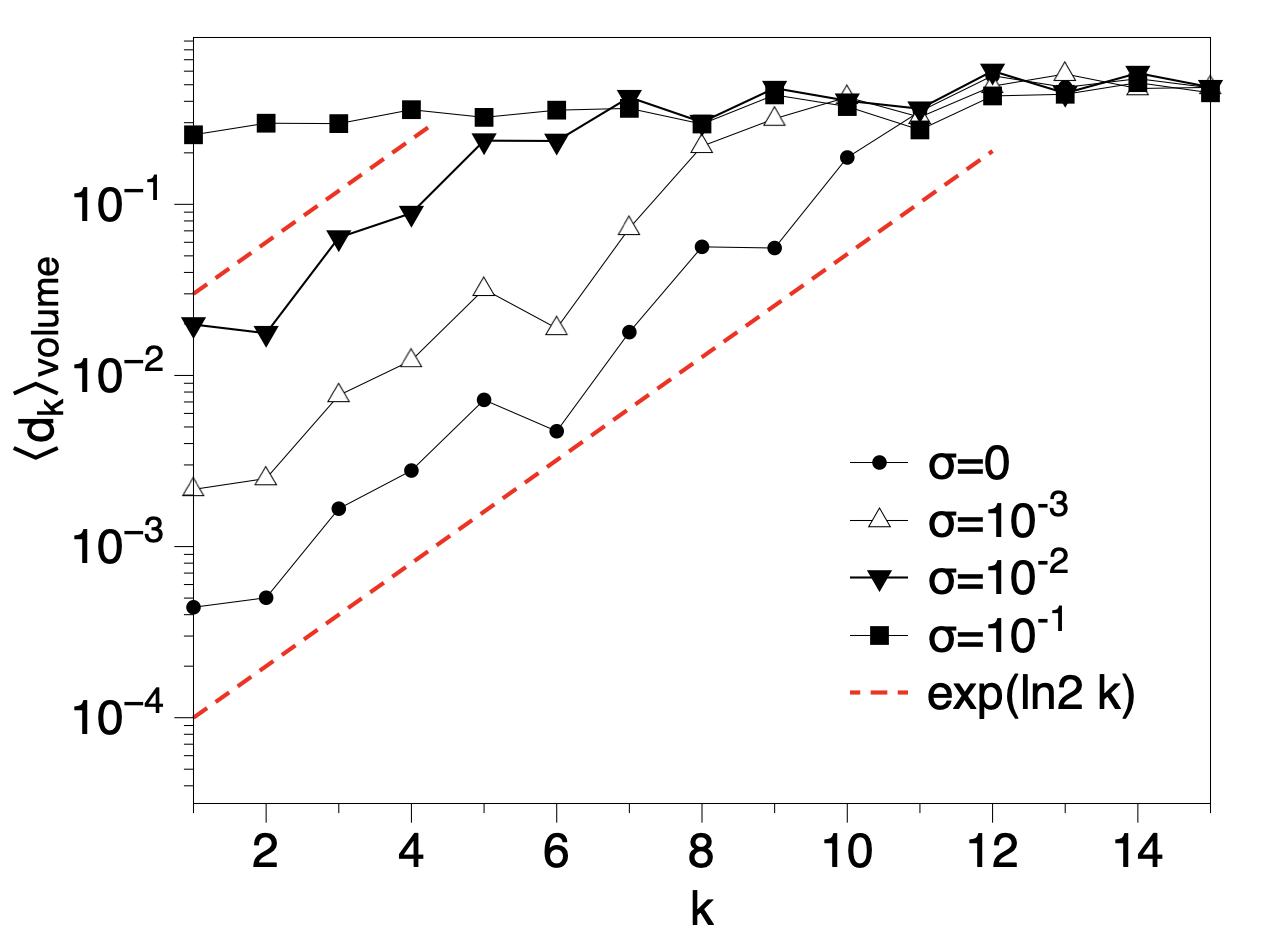}
\par\end{centering}
\caption{Volume-averaged distance $\langle d_k\rangle_{\text{volume}}$  as a function of time $k$, for a network dynamics evolving according to a chaotic logistic map $x_{t+1}=4x_t(1-x_t)$, polluted with extrinsic Gaussian noise ${\cal N}(0,\sigma^2)$ as described in the text, for four different noise intensities $\sigma=0,10^{-3},10^{-2},10^{-1}$.
The exponential expansion phase --which systematically suggests the same exponent $\ln 2$, as expected-- is gradually erased as the noise intensity increases.}
\label{fig:chaos_noise}
\end{figure}

\subsection{Noisy chaotic networks}
To explore how noise contamination can complicate the estimation of the nMLE, we proceed to generate a temporal network $\cal S$ from Eq.~(\ref{eq:log4}) by using the dictionary trick, where before the network mapping, the original chaotic signal is contaminated by a certain amount of white Gaussian noise ${\cal N}(0,\sigma^2)$\footnote{Note that we discard realizations of the noise that take the scalar variable outside the unit interval.}. As we did in Section~\ref{section:iid}, we remove potential algorithmic biases by discarding $\langle d_0 \rangle$ for the computation of $\Lambda$.  Results are summarised in Fig.~\ref{fig:chaos_noise}. The main observation is that noise pollution tends to reduce the extent of the exponential phase (i.e., the saturation time $\tau$ decreases). For small amounts of noise, this phase is still observable, and the estimated nMLE continues to be consistent with  that of the noise-free case. When the noise intensity is above a certain threshold, noise effectively hides the chaotic signal, and the exponential phase can no longer be identified, resulting in an apparent vanishing nMLE. These results are consistent with intuition and with the typical phenomenology observed in noisy chaotic time series \cite{wolf}, \cite{kantz}.


\subsection{Results for the parametric logistic map}
Here we consider  the logistic map $x_{t+1}=rx_t(1-x_t)$. For each value of the parameter $r>r_{\infty}$, using the dictionary trick we generate a long sequence of networks ${\cal S}_r$ with the desired chaoticity properties, and proceed to estimate the network Lyapunov exponent using the method detailed in Section \ref{section:method}. In panel (a) of Fig.\ref{fig:lyap_feig} we plot $\lambda_{\text{nMLE}}^W$ vs $\lambda_{\text{MLE}}$ of the map, for a range of values of the parameter $r$. The agreement is excellent in the region of parameters where the temporal network is chaotic. 


\begin{figure}[htb]
\begin{centering}
\includegraphics[width=0.48\textwidth]{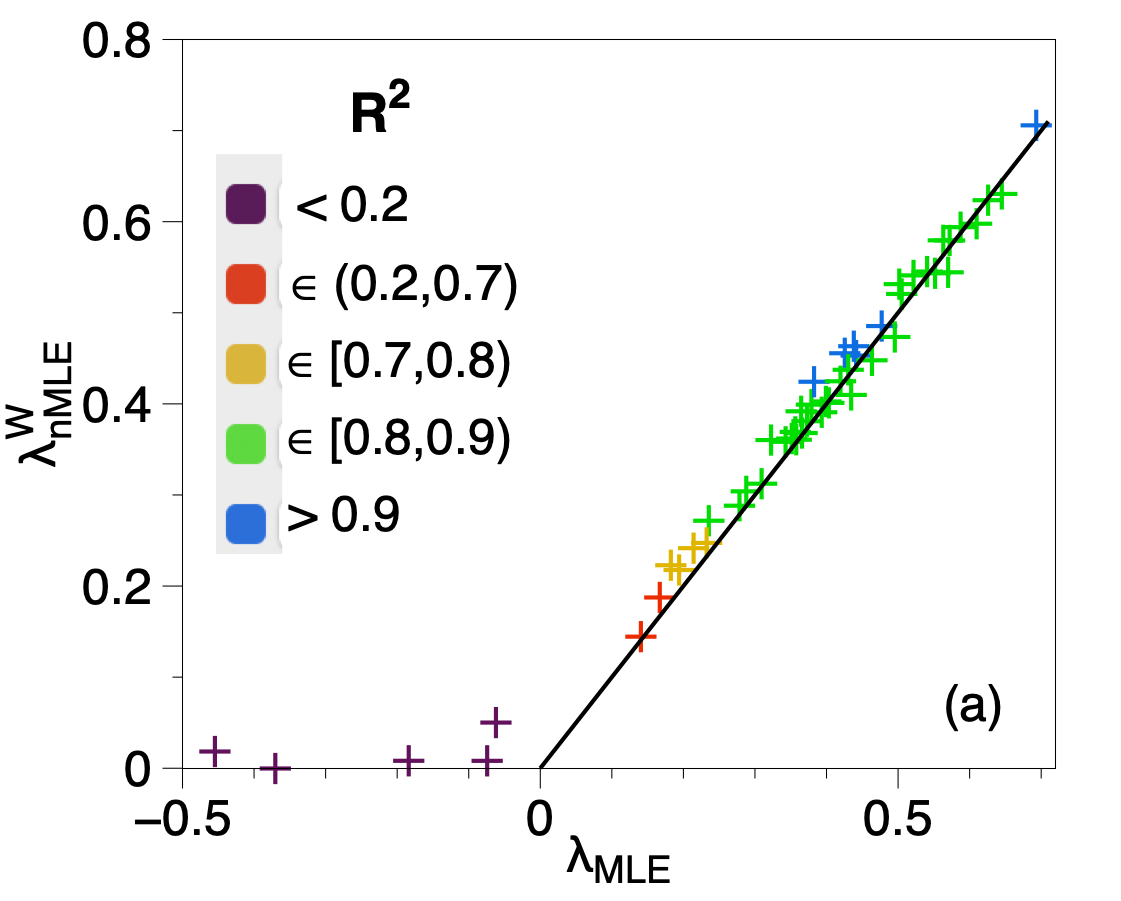}
\includegraphics[width=0.48\textwidth]{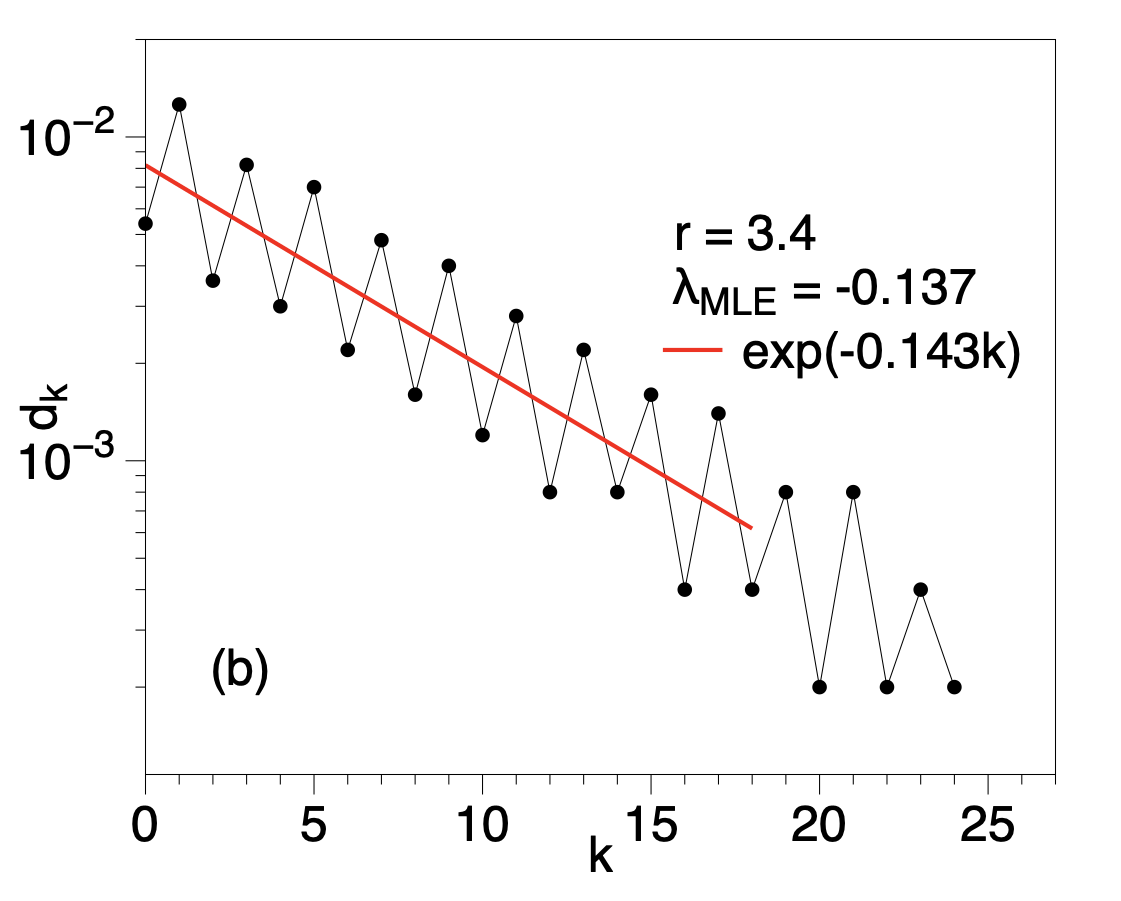}
\par\end{centering}
\caption{
Panel (a): Scatterplot of $\lambda_{\text{nMLE}}^W$, estimated from a temporal network ${\cal S}_r$ generated via the dictionary trick (see the text) from a logistic map $x_{t+1}=rx_t(1-x_t)$ for a range of values of $r$, vs the ground true $\lambda_{\text{MLE}}$. 
The solid line is the diagonal of perfect agreement $y=x$,
highlighting the good agreement found in the chaotic region. The legend states the goodness of fit metric $R^2$ of the fit of $d_k$ to an exponential function. The method is unable to capture negative Lyapunov exponents (observe that in those cases the $R^2$ of the exponential fit is very bad), but these cases can easily be identified as periodic orbits using the autocorrelation function \cite{ACF}, see text for details.  Panel (b): Estimate of the negative nMLE using two initially close temporal networks generated via the dictionary trick from the logistic map at $r=3.4$ (period-2 orbit), where one initial condition belongs to the period-2 attractor and the other is outside the attractor (see the text for details).}
\label{fig:lyap_feig}
\end{figure}

\subsection{A note on negative Lyapunov exponents}
The classical approach to estimate the MLE from a single trajectory displayed by Wolf and Kantz algorithms --based on recurrences of the trajectory-- is, by construction, unable to capture negative MLEs. The reason is straightforward: once in the periodic attractor, the trajectory sequentially visits each element of the periodic orbit, and thus we won't find recurrences that are close but away from the initial condition of interest. Accordingly, our method to estimate nMLE cannot work in that case for the same reasons, as confirmed in Fig.~\ref{fig:lyap_feig}(a). This drawback can be solved using two alternative approaches.\\

 First, it is well known that a periodic time series has an autocorrelation function that peaks at the period of the time series. Interestingly, a recent work \cite{ACF} has operationalised a way to estimate the autocorrelation function of temporal networks, whereby temporal networks that display periodicity are well characterised by a network version of the autocorrelation function. Accordingly, from a practical point of view, before attempting to estimate the nMLE of a given temporal network, it is sensible to apply the procedure of \cite{ACF} and exclude that the temporal network is periodic --which would typically\footnote{Some pathological cases exist for which we can have seemingly periodic behavior but not a negative MLE, e.g. when we have a disconnected attractor composed by a number of bands and a trajectory that periodically visits the different chaotic bands} mean a negative nMLE--. 
 Once this test is done, it is sensible to conduct the nMLE analysis presented in this paper.\\

 Second, it is indeed possible to estimate negative nMLEs if one has access to the latent graph dynamical system (GDS), as in this case one does not need to undergo a Wolf/Kantz approach and one can generate through the GDS temporal networks from close initial graph conditions. To illustrate this, in panel (b) of Fig.~\ref{fig:lyap_feig} we plot the graph distance of two initially close networks evolving according to the logistic map for a value of the map's parameter for which the orbit is periodic (the TNs are again generated via the dictionary trick). One initial condition is set at one of the orbit elements, whereas the other initial condition is a network close in graph space (but outside the periodic attractor). As we can see, there is an exponential shrinking of the initial distance, and the slope gives an estimate of the nMLE, which in this case is negative and in good agreement with the theoretical result.

\section{High-dimensional chaotic networks}
\label{section:validation_highdim}
We now consider the case of high-dimensional chaotic dynamics for temporal networks. We first introduce a generative model, based on coupled Map Lattices (CML). These are high-dimensional dynamical systems with discrete time and continuous
state variables, widely used to model complex spatio-temporal dynamics \cite{kaneko} in disparate contexts such as turbulence
\cite{kanekobook, Hilgers1}, financial markets \cite{Hilgers2}, biological systems \cite{Bevers} or quantum field theories \cite{beck}.\\ 
Globally Coupled Maps (GCMs) \cite{kaneko2} are a mean-field version of CMLS, where the diffusive coupling between the entities in a CML is replaced with an all-to-all coupling, mimicking the effect of a mean-field.
We consider a globally coupled map of $m$ entities, of the form
\begin{equation}
x_i(t+1) = (1-\alpha)F[x_i(t)] + \frac{\alpha}{m}\sum_{j=1}^m F[x_j(t)], \ i=1,2,\dots,m, 
\label{eq:GCM}
\end{equation}
where $F(x)=4x(1-x), \ x\in[0,1]$, where $\alpha \in [0,1]$ is the strength of the mean-field coupling. In the uncoupled case $\alpha=0$, the system is composed of $m$ independent fully chaotic dynamics. Its attractor is thus high-dimensional and,
since there are $m$ Lyapunov exponents all equal to $\ln 2$, we have $\lambda_{\text{MLE}}=\ln 2$.\\
At the other extreme, for complete coupling $\alpha=1$, the system is fully synchronized (i.e., for any time $t$ we have $x_i(t)=x_j(t)$ for all $i,j$), and the dynamics is reduced to the  one-dimensional dynamics, again with $\lambda_{\text{MLE}}=\ln 2$. We add that complete synchronization is in fact known to occur for $\alpha > 1/2$ \cite{kaneko2}.\\
For intermediate coupling the system shows a number of different macroscopic phases \cite{kaneko2}. Among these one finds high-dimensional chaos for weak coupling $\alpha < 0.2$. This is the so-called `turbulent  state'. Interestingly, for CMLs with diffusive coupling, a scaling law has been established \cite{MLE_CML}
\begin{equation}
    \lambda_{\text{MLE}} = \log 2 - \beta {\alpha}^{1/p},
    \label{eq:scaling}
\end{equation}
where $p$ indicates the type of nonlinearity of $F(x)$, i.e. $p=2$ for the logistic map, $p=1$ for tent maps, etc. Results for GCM are less clear. However, when the mean-field coupling can be considered `thermalized' (i.e., independent of x) \cite{GCM_2, GCM_3} then Eq.~(\ref{eq:scaling}) holds for $\beta=1$. However such thermalization is known to be true only for tent maps ($p=1$) and not logistic maps.

Here we consider the range $\alpha \in [0,0.2]$, i.e., the turbulent state of the GCM. We interpret the collection $\{x_i\}_{i=1}^m$ as the (weighted) edge set of a fully connected undirected network backbone of $n$ nodes and $m=n(n-1)/2$ edges. Once the time series of each edge $\{x_i(t)\}_{t=1}^T$ has been computed from Eq.~(\ref{eq:GCM}), we proceed to binarise each edge activity by using a two-symbol generating partition as follows: values $x_i(t)<1/2$ are mapped into the symbol $0$, and $x_i(t)\geq1/2$ onto the symbol $1$\cite{LM}. Note that the use of a generating partition ensures that the symbolised (binary) series preserves the chaotic properties of the original signal \cite{SN1, SN2, SN3}. Finally, we convert the (binary) evolution of the  edges into a time-dependent adjacency matrix, thereby constructing a temporal network $\cal S$. For values of $\alpha$ in the weak-coupling regime, we expect the temporal network to display sensitive dependence on initial conditions.\\  

\begin{figure}[htb]
\begin{centering}
\includegraphics[width=0.48\textwidth]{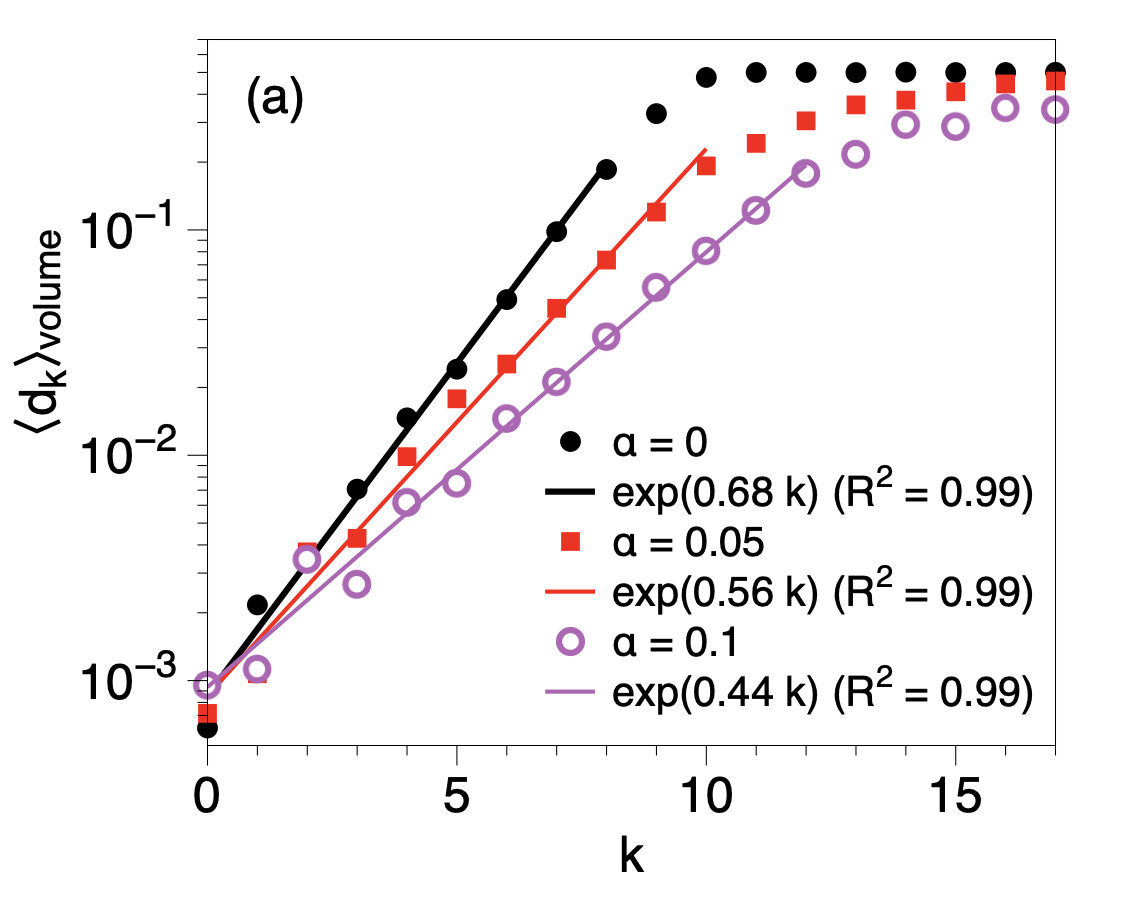}
\includegraphics[width=0.48\textwidth]{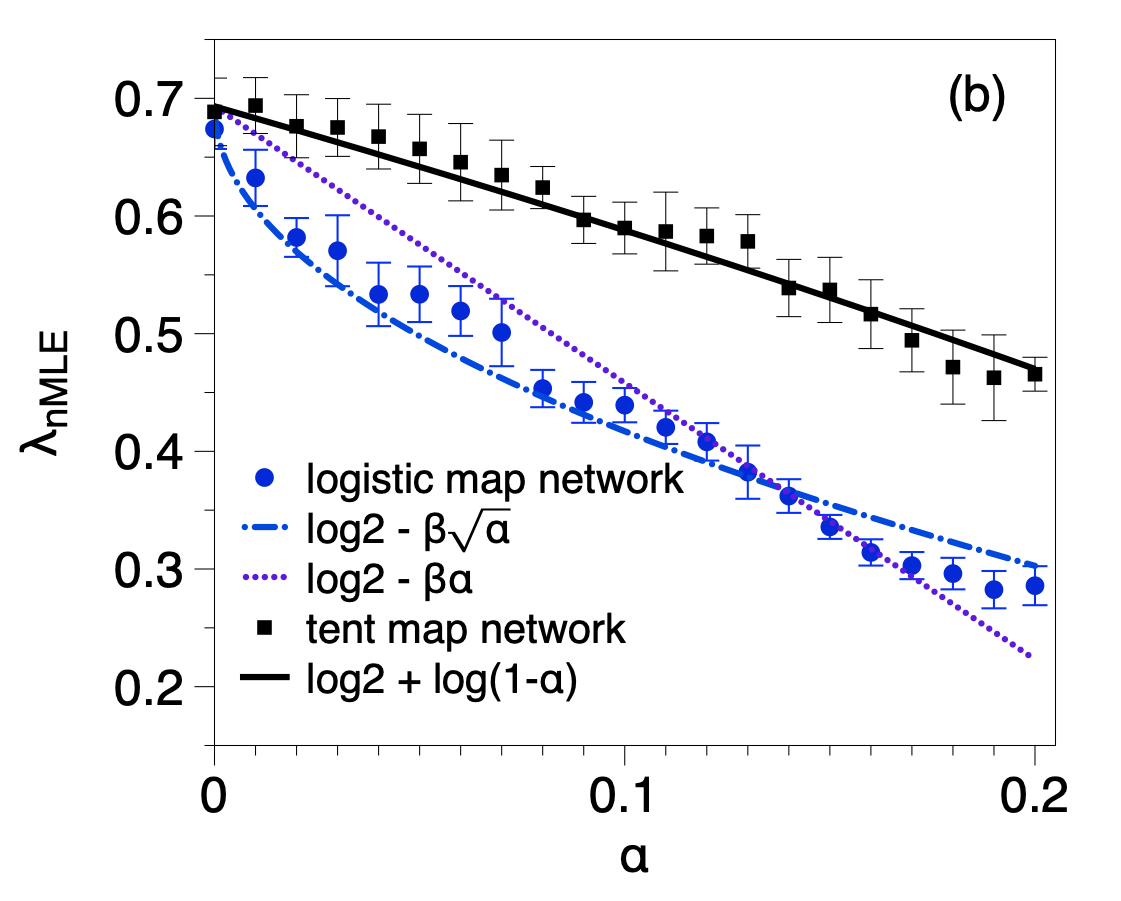}
\par\end{centering}
\caption{Panel (a): Semi-log plot of the volume-averaged distance $\langle d_k\rangle_{\text{volume}}$ as a function of the time step $k$, for a temporal network extracted from the GCM model with coupling constant $\alpha=0,0.05,0.1$. We observe an exponential phase, with different exponents for each value of the coupling constant. The solid lines are the best exponential fits. Panel (b): Estimate of the network maximum Lyapunov exponent $\lambda_{\text{nMLE}}$ vs the coupling constant $\alpha$ for temporal networks generated from a GCM of logistic maps (blue circles) and tent maps (black squares). For each $\alpha$, a total of $50$ initial conditions were considered, and a ball of $100$ points for each initial condition was used. Error bars are standard deviation from the average over 50 different temporal network realizations. Blue lines report the theoretical predictions for logistic and tent CMLs [Eq.~(\ref{eq:scaling})], whereas the black line reports the theoretical prediction for GCM with thermalised mean-field, applicable for tent GCMs only. }
\label{fig:GCM}
\end{figure}

In practice, the Wolf/Kantz methods of inferring the larges Lyapunov exponent proposed in the paper would require a very long sequence $\cal S$ for close enough recurrences to be observable in a system with large $n$. However, here we have access to the actual underlying GDS via Eq.~(\ref{eq:GCM}). Given that the goal of this section is to show evidence that high-dimensional chaotic networks can be generated and their nMLE be estimated, we can use the GDS to generate the temporal network for any required initial condition. Accordingly, for a given initial condition $\{x_i(0)\}_{i=1}^m$, we construct a perturbed copy $\{x_i'(0)\}_{i=1}^m$ (where $\lvert x_i'(0) - x_i(0) \rvert < \epsilon$ for some small choice of $\epsilon$), generate  temporal networks for both of these initial conditions, and track the network distance between the copies over time. We do this for $100$ replicas to extract a volume-averaged distance, and then for $50$ different initial conditions. Observe that, at odds with the model developed in the previous section, here the number of edges in each network snapshot is not fixed, and thus the network phase space is substantially larger. Similarly, the normalization factor of the distance function is now simply the total number of possible edges, $n(n-1)/2$.\\

Results for a network of $n=100$ nodes are shown in Fig.~\ref{fig:GCM}. In panel (a) we plot $\langle d_k\rangle_{\text{volume}}$ vs time $k$, for three different coupling constants $\alpha=0,0.05,0.1$ in the weak coupling regime. In every case we find a clear exponential phase. The exponent in the uncoupled phase $\alpha=0$ is indeed equal to $\ln 2$, as expected, further validating the method. For increasing values of the coupling, interestingly, the nMLE seems to decrease and, as a byproduct, the saturation time $\tau$ increases. In Fig.~\ref{fig:GCM}(b) we plot, as blue dots, the estimated $\lambda_{\text{nMLE}}$ 
as a function of the coupling $\alpha\in [0,0.2]$, indeed showing a clear decrease. Such decrease might be induced by the fact that the $m$ degrees of freedom are now coupled in some nontrivial way.  Blue lines correspond to the theoretical predictions for logistic and tent CMLs obtained from Eq.~(\ref{eq:scaling}). For completeness, we repeated the same analysis for network GCMs constructed from tent maps where $\lambda_{\text{MLE}}$ is explicitly known (black line): $F(x)=1- 2\vert x\vert$, with $x \in [-1,1]$ and a symbolisation partition with $x<0$ mapped to the symbol $0$, and $x \geq 0$ mapped to $1$. Results for this case are plotted as black squares in Fig.~\ref{fig:GCM}(b)\\
We conclude that (i) the TN thereby generated exhibits high-dimensional chaos and its nMLE, reconstructed with the methods we have developed, shows the expected behaviour, and (ii) this validation shows that the method works with TNs where not only the position but also the total number of edges itself fluctuates over time.


\section{Discussion}\label{section:conclusion}

In this work, we propose to look at temporal networks as trajectories of a latent Graph Dynamical System (GDS). This interpretation naturally leads us to explore whether these trajectories can show sensitive dependence on initial conditions, a fingerprint of chaotic behaviour. We have proposed a method to quantify this, and defined and computed the network Maximum Lyapunov Exponent (nMLE) for temporal network. Since the latent GDS is rarely available in practice, our algorithm  exploits the recurrences of the temporal network in graph space. It generalizes the classical approaches of Wolf and Kantz to networks. We have validated the method by generating different synthetic GDS with known ground-truth nMLE.\\

Conceptually speaking, quantifying chaos in the trajectory of structured objects (in our case, mathematical graphs) is somewhat close in spirit to quantifying the dynamical stability of (lattice) spin systems. Thus our approach shares some similarities with the damage-spreading \cite{DS1} and self-overlap methods \cite{SO} in statistical physics, and their applications to cellular automata \cite{CA} and random Boolean systems \cite{lyap_boolean}.

Observe that we have focused on exponential expansion on nearby conditions --i.e., sensitive dependence--, since one of the goals of the paper is to conceptually postulate the existence of chaotic networks and to potentially operationalise a way to measure this deterministic fingerprint in observed TNs, without needs to having access to the underlying GDS. However, our approach can be straightforwardly extended to non-exponential divergence, e.g. algebraic or otherwise, simply by suitably modifying the definition of expansion rates, thus yielding a way to quantify other types of dynamical instability. 

The rationale of this work is to consider graphs evolving over time as whole --yet not punctual-- objects \cite{ACF}, and thus consider its evolution in graph space. It is however true that this approach might have a limitation for (large) real-world temporal networks, as it is often difficult to observe recurrences in high-dimensional space. A possible solution is to extract suitable scalar variables from the network, analyse sensitive dependence on initial conditions in each of them, and extract a consensus. We leave this approach for future work.

Observe that throughout this work we have considered labelled networks. This choice was used for, convenience, illustration, and because TNs are usually labelled, but we expect that a similar approach is possible for unlabelled TNs, i.e. graphs that  evolve over time according to a certain graph dynamics. In this latter case, each network snapshot is no longer uniquely represented by a single adjacency matrix, in the sense that permutations of the rows and columns of the matrix lead to an equally valid description. It is then clear that one needs to use graph distances showing invariance under permutation of rows and columns in the adjacency matrices \cite{AGT}. This could be, for example, distances based on the network spectrum, or graph kernels \cite{kernels}. We leave this interesting extension as a question for further research, as well as the quantification of the full Lyapunov spectrum beyond the maximum one.\\ 

 Finally we would like to add that the fact that the method does not rely on knowing the GDS and instead directly estimates the nMLE from temporal network trajectories enables the investigation of these matters in empirical temporal networks. We foresee a range of potentially interesting applications in physical, biological, economic and social sciences --as indeed temporal networks pervade these disciplines--. This approach is specially appealing in those systems where we don't have access to the `equations of motion' but it is sensible to expect some underlying deterministic dynamics, i.e. physical systems, but the approach is also extensible to systems with socially or biologically-mediated interactions, for instance: do flocks of birds \cite{flock1, flock2, flock3} or crowd behavior \cite{helbing}, adequately modelled as temporal proximity networks, show chaos?


\subsubsection*{Appendix: Graph distances}
Consider two adjacency matrices $A$, and $B$, each with binary entries ($0$ or $1$), describing two simple unweighted graphs with $n$ nodes. The so-called \textit{edit distance} \cite{graph_distance} is a matrix distance defined as
\begin{equation}
    d(A,B) = \sum_{i,j=1}^n \vert a_{ij} - b_{ij} \vert.
\end{equation}
The object $d(A,B)$ counts the number of entries that are different in $A$ and $B$. \\

For simple undirected graphs (symmetric adjacency matrices), we need to account for the fact that the number of edges is only half the number of positive entries of the adjacency matrix, and therefore $d(A,B)/2$ measures the number of edges that exist in one graph but not on the other. We have $d(A,B)/2=0$ if and only if $A=B$. It is also easy to see that $d(A,B)/2$ only takes integer values for symmetric adjacency matrices $A$ and $B$. If $A\ne B$, then $1\le d(A,B)/2\le n(n-1)/2$. We have $d(A,B)/2=1$ when the two graphs are identical except for one edge, which is present in one graph and absent in the other.\\
One can directly use this unnormalized distance (as we do in Section \ref{section:iid}) or 
subsequently normalize $d(A,B)$ using different strategies, e.g. one can divide it over $n(n-1)/2$ (as we do in Section \ref{section:validation_highdim}), or just divide over the maximum possible distance, if further restrictions are imposed between $A$ and $B$ (as we do in Section \ref{section:validation_lowdim}).

If we further impose that both graphs have the same number of edges, then the lower bound cannot be attained and  $2\le d(A,B)/2$ when $A \ne B$. This lower bound is reached when we only need a single edge rewiring to get from the first graph to the second. One can thus define the \textit{rewiring distance}
\begin{equation}
    d(A,B) = \frac{1}{4}\sum_{i,j=1}^n \vert a_{ij} - b_{ij} \vert.
\end{equation}
applicable for simple graphs (i.e. no self-links). This quanity measures the total number of rewirings needed to transform $A$ into $B$ when the associated graphs are simple, unweighted, undirected (symmetric adjacency matrices) and have the same number of nodes and edges.\\

The rewiring distance above is based on the concept of non-overlapping edges, i.e., edges that are present in one graph, but not in the other. Thus, the edit and rewiring distances are based on $\vert a_{ij}-b_{ij}\vert$ for the different edges, and hence assign the same importance to the presence or absence of an edge. One can instead construct measures of distance based on the number of links that are present in both networks. If the edge ${ij}$ is present in both graphs then $a_{ij}b_{ij}=1$, while this product is zero otherwise.  One can prove that the following function is a distance \cite{mason}:
\begin{equation}
    d(A,B) = 1- \frac{1}{2\vert E \vert }\sum_{i,j=1}^n  a_{ij}b_{ij}.
\end{equation}
We replicated the analysis in Sec.~  \ref{section:validation_lowdim} for the distance defined above, and results (resulting nMLE) coincide.\\



\noindent {\bf Acknowledgments} We thank Federico Battiston for helpful comments on initial phases of this research, and Emilio Hernández-Fernández, Sandro Meloni, Lluis Arola-Fernández, Ernesto Estrada, Massimiliano Zanin, Diego Pazó and Juan Manuel López for helpful discussions around several aspects of the work. AC acknowledges funding by the Maria de Maeztu Programme (MDM-2017-0711) and the AEI under the FPI programme. LL acknowledges funding from project DYNDEEP (EUR2021-122007), and LL and VME acknowledge funding from project MISLAND (PID2020-114324GB-C22), both projects funded by the Spanish Ministry of Science and Innovation. This work has been partially supported by the María de Maeztu project CEX2021-001164-M funded by MCIN/AEI/10.13039/501100011033.  \\

\end{document}